\documentclass[12pt,a4paper]{article}
\pdfoutput=1
\usepackage[T1]{fontenc}
\usepackage[utf8]{inputenc}
\usepackage{a4wide}
\usepackage{amsmath}
\usepackage{amssymb}
\usepackage{amsfonts}
\usepackage{cite}
\usepackage[table]{xcolor}
\usepackage{arydshln}
\usepackage{adjustbox}
\usepackage{multirow}
\usepackage{pdflscape}
\usepackage{textcomp}
\usepackage{gensymb}
\usepackage{graphicx}
\graphicspath{{./figs/}}
\usepackage{diagbox}
\usepackage{pifont}
\usepackage{bigstrut}
\usepackage[small,bf]{caption}
\usepackage{tabulary}
\usepackage{booktabs}
\usepackage{stackengine}
\usepackage{mathtools}
\usepackage{enumerate}
\usepackage[bb=boondox]{mathalfa}
\usepackage[
colorlinks=true,
linkcolor=red!50!black,
citecolor=teal,
urlcolor=blue]{hyperref}
\usepackage{mathrsfs}
\usepackage{slashed}
\usepackage{cancel}
\usepackage{listings}
\usepackage{fontawesome5}
\usepackage{seqsplit}

\def\eps{\epsilon}

\def\1{\mathbf{1}}

\def\2{\mathbf{2}}
\def\3{\mathbf{3}}

\DeclarePairedDelimiter{\abs}{\lvert}{\rvert}

\definecolor{darkgreen}{rgb}{0.0, 0.5, 0.0}
\allowdisplaybreaks
\makeatletter
\g@addto@macro\bfseries{\boldmath}
\let\oldabs\abs
\def\abs{\@ifstar{\oldabs}{\oldabs*}}
\makeatother

\begin{document}



\thispagestyle{empty}
\renewcommand*{\thefootnote}{\fnsymbol{footnote}}

\begin{center}
{\bf\Large \textsc{FeynRules} and \textsc{UFO} models for $\nu$SMEFT: \\[2mm]
operators of dimensions five and six}\\[8mm]

\textbf{Arsenii~Titov}$^{\,a,\,b,\,}$%
\footnote{\href{mailto:arsenii.titov@unipd.it}{\texttt{arsenii.titov@unipd.it}}}
\\[4mm]
\textit{$^{a}$Dipartimento di Fisica e Astronomia ``Galileo~Galilei'', \\
Università degli Studi di Padova,  
Via Francesco~Marzolo 8, 35131 Padova, Italy} \\[2mm]
\textit{$^{b}$INFN, Sezione di Padova,  
Via Francesco~Marzolo 8, 35131 Padova, Italy}
\end{center}
\vspace{4mm}

\begin{abstract}
\noindent 
We release \textsc{FeynRules} and \textsc{UFO} model files 
for the $\nu$SMEFT~---~the effective field theory of the Standard Model extended with right-handed neutrinos, $N_R$.
These model files include  
operators with $N_R$ up to dimension six. 
In particular, we provide a \textsc{UFO} implementation of four-fermion 
operators with Majorana $N$ compatible with \textsc{MadGraph5\_aMC@NLO}. 
It relies on introducing auxiliary, non-propagating fields.
All the model files are made publicly available 
on \href{https://github.com/arsenii-titov/vSMEFT.git}{\textsc{GitHub}~\faGithub}.
\end{abstract}
\renewcommand*{\thefootnote}{\arabic{footnote}}
\setcounter{footnote}{0}
\vspace{8mm}

\clearpage
\tableofcontents

\section{\label{sec:intro}Introduction}
The existence of right-handed (RH) neutrinos, $N_R$, is primarily 
motivated by non-zero masses of light, standard-model-like neutrinos. 
Adding the Yukawa interaction of the form $y_\nu \overline{L} \tilde{H} N_R$ 
to the Standard Model (SM) Lagrangian leads to Dirac neutrino masses.
Theoretical issue with this minimal, renormalisable option is that 
the neutrino Yukawa coupling, $y_\nu$, must be tiny 
to account for the observed smallness of neutrino mass, $m_\nu$, namely, 
$y_\nu \sim \mathcal{O}(10^{-13})$ assuming $m_\nu \approx 0.05$~eV. 
Since $N_R$ are complete singlets under the SM gauge group, 
the Majorana mass term $m_N \overline{N_R^c} N_R$ is allowed.
The combination of the Dirac and Majorana mass terms leads 
to the renowned seesaw type I mechanism~\cite{Minkowski:1977sc,Yanagida:1979as,Gell-Mann:1979vob,Glashow:1979nm,Mohapatra:1979ia} 
able to explain the smallness of light neutrino masses 
by a large value of $m_N \sim \mathcal{O}(10^{15})$~GeV for $y_\nu \sim \mathcal{O}(1)$ 
through the seesaw relation: 
\begin{equation}
 m_\nu = - \frac{v^2}{2} y_\nu m_N^{-1} y_\nu^T\,.
\end{equation}
However, if $y_\nu \lesssim y_e \sim 10^{-6}$, the heavy neutrino mass 
$m_N$ is around or below the electroweak scale, set by the Higgs 
vacuum expectation value (VEV), $v \approx 246$~GeV. 
This is entirely plausible possibility, even if less attractive from a purely theoretical standpoint. 

Several open problems/puzzles in particle physics, 
including the hierarchy problem, the flavour problem, the strong CP problem, 
to name several, 
motivate the existence of new physics 
at the mass scale $\Lambda \gg v$.
Assuming that (i) the low-energy spectrum contains, 
in addition to SM degrees of freedom, only $N_R$, 
and (ii) new physics exists at the scale $\Lambda$, 
the most general description of new physics effects at energies 
$E \lesssim \Lambda$ is captured by the effective field theory (EFT) 
of the SM extended with $N_R$, denoted as $\nu$SMEFT 
(or NSMEFT, SMNEFT, $N_R$SMEFT)~\cite{delAguila:2008ir,Aparici:2009fh,Bhattacharya:2015vja,Liao:2016qyd} (see also Refs.~\cite{Bell:2005kz,Graesser:2007yj,Graesser:2007pc} for earlier works that considered some of the operators with $N_R$).
The phenomenology of this EFT has been extensively studied 
over the last years, see \textit{e.g.}~\cite{Duarte:2015iba,Duarte:2016caz,Caputo:2017pit,Bischer:2019ttk,Alcaide:2019pnf,Butterworth:2019iff,Jones-Perez:2019plk,Han:2020pff,Li:2020lba,Biekotter:2020tbd,Li:2020wxi,DeVries:2020jbs,Barducci:2020icf,Cirigliano:2021peb,Cottin:2021lzz,Beltran:2021hpq,Zhou:2021ylt,Zapata:2022qwo,Barducci:2022hll,Delgado:2022fea,Mitra:2022nri,Fernandez-Martinez:2023phj,Duarte:2023tdw,Fuyuto:2024oii,Colangelo:2024sbf,Mitra:2024ebr,Borah:2024twm,Biswas:2024gtr,Duarte:2025zrg,Beltran:2025ilg,Bolton:2025tqw,Alonso:2025gzl}.
However, no \textsc{FeynRules}~\cite{Christensen:2008py,Alloul:2013bka} and \textsc{UFO}~\cite{Degrande:2011ua} model files 
enabling further studies of collider phenomenology of the $\nu$SMEFT  
have been made publicly available, 
except for Ref.~\cite{Cirigliano:2021peb} that added 
one dimension-six operator to the \href{https://feynrules.irmp.ucl.ac.be/wiki/HeavyN}{\texttt{HeavyN}} model~\cite{Alva:2014gxa,Degrande:2016aje}.

We fill this gap by releasing the model files 
including dimension-five and dimension-six operators with $N_R$.
More specifically, we extend the default 
\href{https://feynrules.irmp.ucl.ac.be/wiki/StandardModel}{\texttt{SM}} 
model file 
supplied with \textsc{FeynRules}
and the \href{https://feynrules.irmp.ucl.ac.be/wiki/HeavyN}{\texttt{HeavyN}} model file by adding the  
two-fermion and four-fermion operators with $N_R$.
The resulting \textsc{FeynRules} model files 
\texttt{vSMEFT\_2-fermions\_Higgs-N.fr}, 
\texttt{vSMEFT\_4-fermions\_single-N.fr} 
and \texttt{vSMEFT\_4-fermions\_pair-N.fr},
are built upon those originally developed
in Refs.~\cite{Butterworth:2019iff}, \cite{Beltran:2021hpq} and \cite{Cottin:2021lzz}, 
respectively. 
While for the two-fermion effective interactions the generated 
\textsc{UFO} model 
\texttt{vSMEFT\_2-fermions\_Higgs-N\_UFO}
can be readily used for performing simulations with \textsc{MadGraph5\_aMC@NLO}~\cite{Alwall:2014hca},
the four-fermion interactions with a Majorana $N$ directly 
implemented in \textsc{FeynRules} and exported to \textsc{UFO} 
are not supported by \textsc{MadGraph5\_aMC@NLO}. 
To overcome this issue, in \texttt{vSMEFT\_4-fermions\_single-N\_aux.fr} and 
\texttt{vSMEFT\_4-fermions\_pair-N\_aux.fr}, we provide an implementation of the four-fermion interactions making use of auxiliary, non-propagating scalar and vector fields. 
The resulting UFO models \linebreak \texttt{vSMEFT\_4-fermions\_single-N\_aux\_UFO} 
and \texttt{vSMEFT\_4-fermions\_pair-N\_aux\_UFO} are compatible with \textsc{MadGraph5\_aMC@NLO}. Finally, we combine all the effective interactions 
of interest in a single UFO model \texttt{vSMEFT\_UFO}.
We make all the model files publicly available 
on \href{https://github.com/arsenii-titov/vSMEFT.git}{\textsc{GitHub}~\faGithub}.

The remainder of this note is organised as follows. 
In Sec.~\ref{sec:effops}, we summarise the effective interactions of interest. 
In Sec.~\ref{sec:model-file}, we provide description of the 
released \textsc{FeynRules} and \textsc{UFO} model files, 
detailing the auxiliary-field implementation of four-fermion operators 
with a Majorana $N$ in Sec.~\ref{sec:aux}.
Section~\ref{sec:validation} describes model validation. 
Finally, in Sec.~\ref{sec:outlook}, we provide a brief summary and an outlook.

\section{\label{sec:effops}Effective operators of dimensions five and six}
The renormalisable Lagrangian is given by
\begin{equation}
\mathcal{L}_4 = \mathcal{L}_\mathrm{SM} 
+ \overline{N_{jR}}\, i\slashed{\partial} N_{jR}
- \left[y_\nu^{ij} \overline{L_i} \tilde{H} N_{jR} 
+ \frac{1}{2} m_N^{jk} \overline{N_{jR}^c} N_{kR} + \text{h.c.}\right],
\label{eq:Ldim4}
\end{equation}
%
where $\mathcal{L}_\mathrm{SM}$ stands for the SM Lagrangian, 
$y_\nu^{ij}$ are the neutrino Yukawa couplings, 
with $i= e,\mu,\tau~(1,2,3)$ and $j=1,\dots,n_s$, 
and $m_N^{jk} = m_N^{kj}$ is the Majorana mass matrix of RH neutrinos. 
The sum over repeated flavour indices is assumed.
As usual, 
$L_i= (\nu_{iL},e_{iL})^T$ and $H = (H^+,H^0)^T$ are SU(2)$_L$ doublets,
$\tilde{H} = i \sigma_2 H^\ast$, and 
$N_{jR}^c = C\overline{N_{jR}}^T$, with $C$ being the charge conjugation matrix.
In what follows, we work in the diagonal basis for RH neutrinos, such that 
$m_N^{jk} = m_{N_j} \delta^{jk}$.

\subsection{\label{sec:2fermions}Two-fermion operators with $N_R$}
We extend the renormalisable Lagrangian in Eq.~\eqref{eq:Ldim4} 
with the dimension-five and dimension-six  
two-fermion operators with $N_R$ summarised in Tab.~\ref{tab:HiggsN}.
\begin{table}[t]
 \renewcommand{\arraystretch}{1.75}
 \centering
 \begin{tabular}{|c|c|c|c|c|}
 \hline
 Operator & Structure & $\#_{n_s=1}$ & $\#_{n_s=3}$ & Dimension \\
 \hline 
 $\mathcal{O}_{NNH}^{jk}$ & $(\overline{N_{jR}^c}N_{kR}) (H^\dagger H)$ & 2 & 12 & 5 \\
 $\mathcal{O}_{NNB}^{jk}$ & $\overline{N_{jR}^c} \sigma^{\mu\nu}N_{kR} B_{\mu\nu}$ & 0 & 6 & 5 \\
 \hline
 $\mathcal{O}_{HN}^{jk}$ & $(\overline{N_{jR}} \gamma^\mu N_{kR})(H^\dagger i \overleftrightarrow{D_{\mu}} H)$ & 1 & 9 & 6 \\
 $\mathcal{O}_{HNe}^{ji}$ & $(\overline{N_{jR}} \gamma^\mu e_{iR}) (\tilde{H}^\dagger i D_\mu H)$ & 6 & 18 & 6 \\
 $\mathcal{O}_{LNH}^{ij}$ & $(\overline{L_i} \tilde{H} N_{jR}) (H^\dagger H)$ & 6 & 18 & 6 \\
 $\mathcal{O}_{NB}^{ij}$ & $\overline{L_i} \sigma^{\mu\nu} N_{jR} \tilde{H} B_{\mu\nu}$ & 6 & 18 & 6 \\
 $\mathcal{O}_{NW}^{ij}$ & $\overline{L_i} \sigma^{\mu\nu} N_{jR}\, \sigma^I \tilde{H}\, W^I_{\mu\nu}$ & 6 & 18 & 6 \\
 \hline
 \end{tabular}
 \caption{\label{tab:HiggsN}Dimension-five and dimension-six two-fermion operators with $N_R$. The indices $j,k=1,\dots,n_s$ label RH neutrino generations, whereas the index $i= e,\mu,\tau~(1,2,3)$ stands for the SM lepton flavour. The third (fourth) column shows the number of independent real parameters associated with each operator structure for $n_s=1$ ($n_s=3$).}
\end{table} 
In this table,
\begin{equation}
 H^\dagger \overleftrightarrow{D_\mu} H = H^\dagger D_\mu H - (D_\mu H)^\dagger H\,,
\end{equation}
with the covariant derivative $D_\mu$ following the \textsc{FeynRules} convention; 
$\sigma^{\mu\nu} = i [\gamma^\mu,\gamma^{\nu}]/2$, 
with $\gamma^\mu$, $\mu=0,1,2,3$, being the Dirac matrices;
$\sigma^I$, $I=1,2,3$, are the Pauli matrices.
The dimension-five dipole operator $\mathcal{O}_{NNB}^{jk}$ vanishes identically 
for $n_s=1$, since it is antisymmetric in the RH neutrino generation indices $j$ and $k$.

The Lagrangians describing higher-dimensional Higgs/gauge-$N_R$ interactions at dimensions five and six are
\begin{align}
 \mathcal{L}_5 &= c_{NNH}^{jk}\mathcal{O}_{NNH}^{jk} + c_{NNB}^{jk}\mathcal{O}_{NNB}^{jk} + \text{h.c.}\,,
 \label{eq:L5}\\[0.2cm]
 \mathcal{L}_6^{\text{Higgs-}N_R} &= c_{HN}^{jk}\mathcal{O}_{HN}^{jk} + 
 \left[c_{HNe}^{ji} \mathcal{O}_{HNe}^{ji} 
 + c_{LNH}^{ij}\mathcal{O}_{LNH}^{ij} 
 + c_{NB}^{ij}\mathcal{O}_{NB}^{ij} 
 + c_{NW}^{ij}\mathcal{O}_{NW}^{ij} + \text{h.c.}\right]\,,
 \label{eq:L6}
\end{align}
where $c_{NNH}^{jk}$, $c_{HN}^{jk}$, etc., 
denote dimensionless Wilson coefficients (WCs). 
The operator $\mathcal{O}_{HN} = c_{HN}^{jk} \mathcal{O}_{HN}^{jk}$ is hermitian, whereas all the others are not.

Sometimes it proves convenient to rewrite 
the combination of the dipole operators 
$\mathcal{O}_{NB}$ and $\mathcal{O}_{NW}$
in Eq.~\eqref{eq:L6} as
\begin{equation}
c_{NB}^{ij} \mathcal{O}_{NB}^{ij} + c_{NW}^{ij} \mathcal{O}_{NW}^{ij} 
= c_{NA}^{ij} \mathcal{O}_{NA}^{ij} + c_{NZ}^{ij} \mathcal{O}_{NZ}^{ij}\,,
\end{equation}
where the operators $\mathcal{O}_{NA}^{ij}$ and $\mathcal{O}_{NZ}^{ij}$ 
are the following linear combinations of $\mathcal{O}_{NB}^{ij}$ and $\mathcal{O}_{NW}^{ij}$:
\begin{equation}
 \begin{cases}
 \mathcal{O}_{NA}^{ij} =\phantom{-} c_w \mathcal{O}_{NB}^{ij} + s_w \mathcal{O}_{NW}^{ij}\\
\mathcal{O}_{NZ}^{ij} = -s_w \mathcal{O}_{NB}^{ij} + c_w \mathcal{O}_{NW}^{ij}
 \end{cases},
\end{equation}
and the corresponding WCs are
\begin{equation}
 \begin{cases}
 c_{NA}^{ij} = \phantom{-} c_w c_{NB}^{ij}  + s_w c_{NW}^{ij}  \\
 c_{NZ}^{ij} = - s_w c_{NB}^{ij} + c_w c_{NW}^{ij}  
 \end{cases}.
 \label{eq:cNAcNZ}
\end{equation}
Here $c_w \equiv \cos\theta_w$ and $s_w \equiv \sin\theta_w$, 
with $\theta_w$ being the weak mixing angle.

\subsection{\label{sec:4fermions}Four-fermion operators with $N_R$}
In addition to the two-fermion effective interactions, 
at dimension six there are also four-fermion operators with $N_R$. 
They can be divided into \textit{single-$N_R$} and \textit{pair-$N_R$} operators, 
see Tabs.~\ref{tab:singleN} and \ref{tab:pairN}, respectively.
\begin{table}[t]  
 \renewcommand{\arraystretch}{1.75}
 \centering
 \begin{tabular}[t]{|c|c|c|c|}
  \hline
  Operator & Structure & $\#_{n_s=1}$ & $\#_{n_s=3}$ \\ 
\hline
  $\mathcal{O}_{LNLe}^{ijkl}$ &
  $(\overline{L_i}N_{jR})\, \epsilon\, (\overline{L_k}e_{lR})$ &
  54 & 162 \\
  \hline
  $\mathcal{O}_{duNe}^{ijkl}$ &
  $(\overline{d_{iR}}\gamma^{\mu}u_{jR}) (\overline{N_{kR}}\gamma_{\mu}e_{lR})$ &
  54 & 162 \\ 
 $\mathcal{O}_{LNQd}^{ijkl}$ &
 $(\overline{L_i}N_{jR})\, \epsilon\, (\overline{Q_k}d_{lR})$ &
  54 & 162 \\
  $\mathcal{O}_{LdQN}^{ijkl}$ & 
  $(\overline{L_i}d_{jR})\, \epsilon\, (\overline{Q_k}N_{lR})$ &
  54 & 162 \\
 $\mathcal{O}_{QuNL}^{ijkl}$ &
 $(\overline{Q_i}u_{jR}) (\overline{N_{kR}}L_l)$ &
  54 & 162 \\  
\hline
 \end{tabular}
\caption{\label{tab:singleN}\textit{Single-$N_R$} four-fermion operators. The SM lepton and quark flavour indices run from 1 to 3, 
whereas the RH neutrino generation index runs from 1 to $n_s$. The last two columns show the number of independent real parameters associated with each operator structure for $n_s=1$ and $n_s=3$.}
\end{table}
%
%
\begin{table}[h!]  
 \renewcommand{\arraystretch}{1.75}
 \centering
 \begin{tabular}[t]{|c|c|c|c|}
  \hline
  Operator & Structure & $\#_{n_s=1}$ & $\#_{n_s=3}$ \\ 
  \hline
  $\mathcal{O}_{NNNN}^{ijkl}$ &
  $(\overline{N_{iR}^c} N_{jR}) (\overline{N_{kR}^c} N_{lR})$ &
  0 & 12 \\
  $\mathcal{O}_{NN}^{ijkl}$ &
  $(\overline{N_{iR}}\gamma^{\mu}N_{jR}) (\overline{N_{kR}}\gamma_{\mu}N_{lR})$ &
  1 & 36 \\
  \hline
  $\mathcal{O}_{eN}^{ijkl}$ &
  $(\overline{e_{iR}}\gamma^{\mu}e_{jR}) (\overline{N_{kR}}\gamma_{\mu}N_{lR})$ &
  9 & 81 \\
  $\mathcal{O}_{LN}^{ijkl}$ &
  $(\overline{L_i}\gamma^{\mu}L_j) (\overline{N_{kR}}\gamma_{\mu}N_{lR})$ &
  9 & 81 \\  
  \hline
  $\mathcal{O}_{uN}^{ijkl}$ &
  $(\overline{u_{iR}}\gamma^{\mu}u_{jR}) (\overline{N_{kR}}\gamma_{\mu}N_{lR})$ &
  9 & 81 \\
  $\mathcal{O}_{dN}^{ijkl}$ &
  $(\overline{d_{iR}}\gamma^{\mu}d_{jR}) (\overline{N_{kR}}\gamma_{\mu}N_{lR})$ &
  9 & 81 \\
  $\mathcal{O}_{QN}^{ijkl}$ &
  $(\overline{Q_i}\gamma^{\mu}Q_j) (\overline{N_{kR}}\gamma_{\mu}N_{lR})$ &
  9 & 81 \\   
  \hline
 \end{tabular}
\caption{\label{tab:pairN}\textit{Pair-$N_R$} four-fermion operators. The SM lepton and quark flavour indices run from 1 to 3, 
whereas the RH neutrino generation index runs from 1 to $n_s$. The last two columns show the number of independent real parameters associated with each operator structure for $n_s=1$ and $n_s=3$.}
\end{table} 
In Tab.~\ref{tab:pairN}, we have also included the lepton-number-violating operator 
$\mathcal{O}_{NNNN}$ and the lepton-number-conserving operator $\mathcal{O}_{NN}$, both containing four instances of the RH neutrino field. 
Note that $\mathcal{O}_{NNNN}$ vanishes identically for $n_s=1$.

The corresponding dimension-six effective Lagrangians are
\begin{align}
 \mathcal{L}_6^{\text{single-}N_R} &= c_{LNLe}^{ijkl} \mathcal{O}_{LNLe}^{ijkl} 
 + c_{duNe}^{ijkl} \mathcal{O}_{duNe}^{ijkl} + c_{LNQd}^{ijkl} \mathcal{O}_{LNQd}^{ijkl} 
 + c_{LdQN}^{ijkl} \mathcal{O}_{LdQN}^{ijkl} \nonumber \\
 &\phantom{{}={}}+ c_{QuNL}^{ijkl} \mathcal{O}_{QuNL}^{ijkl} + \text{h.c.}\,, \\[0.2cm]
 \mathcal{L}_6^{\text{pair-}N_R} &= c_{NN}^{ijkl} \mathcal{O}_{NN}^{ijkl} + c_{eN}^{ijkl} \mathcal{O}_{eN}^{ijkl} 
 + c_{LN}^{ijkl} \mathcal{O}_{LN}^{ijkl} 
 + c_{uN}^{ijkl} \mathcal{O}_{uN}^{ijkl} + c_{dN}^{ijkl} \mathcal{O}_{dN}^{ijkl} + c_{QN}^{ijkl} \mathcal{O}_{QN}^{ijkl} \nonumber \\
 &+\left[c_{NNNN}^{ijkl} \mathcal{O}_{NNNN}^{ijkl} + \text{h.c.}\right]\,.
\end{align}
The full Lagrangian describing $N_R$ interactions up to dimension six is given by%
\footnote{For simplicity, in what follows, 
we assume the same scale $\Lambda$ 
for the lepton-number-violating 
and the lepton-number-conserving operators.}
\begin{equation}
 \mathcal{L} = \mathcal{L}_4 + \frac{1}{\Lambda}\mathcal{L}_5 
 + \frac{1}{\Lambda^2} \left(\mathcal{L}_6^{\text{Higgs-}N_R} 
 + \mathcal{L}_6^{\text{single-}N_R} + \mathcal{L}_6^{\text{pair-}N_R}\right)\,.
 \label{eq:L}
\end{equation}
As can be seen from the tables, the number of parameters associated with 
the four-fermion operators proliferates significantly with $n_s$. 
In our numerical implementation, we focus on the minimal case of $n_s=1$, 
\textit{i.e.} we assume effective couplings to the first generation of RH neutrinos only.

\section{\label{sec:model-file}Model implementation}
%
%
\subsection{\label{sec:FR}\textsc{FeynRules} models}
%
In this section, we describe the structure and the main features of the 
\textsc{FeynRules} model files 
\texttt{vSM.fr}, \texttt{vSMEFT\_2-fermions\_Higgs-N.fr}, 
\texttt{vSMEFT\_4-fermions\_single-N.fr} and \texttt{vSMEFT\_4-fermions\_pair-N.fr}
containing the renormalisable interactions ($\mathcal{L}_4$), 
higher-dimensional two-fermion interactions with $N_R$ ($\mathcal{L}_5$ and $\mathcal{L}_6^{\text{Higgs-}N_R}$), four-fermion single-$N_R$ ($\mathcal{L}_6^{\text{single-}N_R}$) and pair-$N_R$ ($\mathcal{L}_6^{\text{pair-}N_R}$) interactions, respectively. 
Such a modular structure allows one to work with a given class of effective operators of interest. Finally, \texttt{vSMEFT.fr} combines all the operators in a single model.

\subsubsection{\label{sec:vSM}\texttt{vSM.fr}}
%
Following the \href{https://feynrules.irmp.ucl.ac.be/wiki/HeavyN}{\texttt{HeavyN}}  model file, 
massive Majorana neutrinos $N_j = N_j^c$, $j=1,\dots,n_s$, 
also known as heavy neutral leptons (HNLs),
are defined in \texttt{vSM.fr} as follows:
\begin{lstlisting}
(* HNL: physical fields *)
F[131] == {
 ClassName       -> N1,
 SelfConjugate   -> True,
 Mass            -> {mN1,300.},
 Width           -> {WN1,0.303},
 PropagatorLabel -> "N1",
 PropagatorType  -> Straight,
 PropagatorArrow -> False,
 ParticleName    -> "N1",
 PDG             -> {9900012},
 FullName        -> "N1"
}
\end{lstlisting}
We define their RH components as unphysical fields:
\begin{lstlisting}
(* HNL: unphysical fields *)
F[141] == {
 ClassName      -> NR,
 Unphysical     -> True,
 Indices        -> {Index[HNLgeneration]},
 FlavorIndex    -> HNLgeneration,
 SelfConjugate  -> False,
 QuantumNumbers -> {Y->0, LeptonNumber->1},
 Definitions    -> {NR[sp1_,gg_] :> Module[{sp2}, ProjP[sp1,sp2] N1[sp2]]}
}
\end{lstlisting}
Here, the newly introduced index \texttt{HNLgeneration}, 
corresponding to $n_s$, is fixed to 1 for simplicity, 
\textit{i.e.}~we assume non-renormalisable couplings to the first generation of RH neutrinos only.
A generalisation to the case of $n_s = 3$ (or more) heavy neutrinos is straightforward.

We also add the one-loop Higgs-gluon-gluon vertex 
as an effective coupling in the large top mass limit, 
similarly to how it is implemented in the 
\href{https://smeftsim.github.io}{\texttt{SMEFTsim}} package~\cite{Brivio:2017btx,Brivio:2020onw}:%
\footnote{We note that \texttt{SMEFTsim 3.0}~\cite{Brivio:2020onw} 
has a more refined implementation of loop-induced Higgs couplings to gauge bosons.}
\begin{lstlisting}
(* Effective 1-loop GGh coupling *)
Ifermion[x_,y_]:= 1/3 + (11 y)/90 + (22 y^2)/315 + (74 y^3)/1575 + (7 x)/90 + (16 y x)/315 + (58 y^2 x)/1575 + (2 x^2)/63 + (2 y x^2)/75 + (26 x^3)/1575;

LSMloop := Block[{mu,nu,aa},
    gs^2/(16 Pi^2) Ifermion[MH^2/(4 MT^2),0] (H/vev) (del[G[nu,aa],mu] - del[G[mu,aa], nu])^2
]
\end{lstlisting}
This allows one to simulate, in particular, the SM 
Higgs production via gluon fusion.

\subsubsection{\label{sec:vSMEFTHiggsN}\texttt{vSMEFT\_2-fermions\_Higgs-N.fr}}
%
Using the defined class \texttt{NR} and the unphysical fields 
declared in the SM model file \texttt{\seqsplit{SM\_v\_Majorana.fr}},%
\footnote{In \texttt{SM\_v\_Majorana.fr}, we set 
neutrinos \texttt{vl} to be Majorana, \textit{i.e.} \texttt{SelfConjugate -> True}, 
changing also \texttt{PropagatorArrow -> False} and commenting out
\texttt{QuantumNumbers -> \{LeptonNumber -> 1\}} and 
\texttt{AntiParticleName -> \{"ve\textasciitilde","vm\textasciitilde","vt\textasciitilde"\}}.
These are the only differences with respect to the default 
\href{https://feynrules.irmp.ucl.ac.be/wiki/StandardModel}{\texttt{SM}} model file $\texttt{SM.fr}$ supplied with \textsc{FeynRules}.}
we implement in \texttt{vSMEFT\_2-fermions\_Higgs-N.fr} 
the higher-dimensional operators given in Tab.~\ref{tab:HiggsN}. 
For instance, for the $d=5$ operator 
$\mathcal{O}_{NNH} = c_{NNH}^{11}\mathcal{O}_{NNH}^{11}$,
we define
\begin{lstlisting}
(* Wilson coefficients cNNH *)
cNNH == {
 ParameterType	  -> External,
 BlockName	  -> OPERATORNNH,
 Indices	  -> {Index[HNLgeneration], Index[HNLgeneration]},
 ComplexParameter -> False,
 Value            -> {cNNH[i_?NumericQ, j_?NumericQ] -> 0.0},
 InteractionOrder -> {NP, 1},
 TeX		  -> Subscript[c, NNH],
 Description      -> "Wilson coefficients cNNH"
}
\end{lstlisting}
and
\begin{lstlisting}
(* Operator ONNH *)
ONNH := Block[{opNNH,sp1,ii,gg1,gg2,feynmangaugerules},
 feynmangaugerules = If[Not[FeynmanGauge], {G0|GP|GPbar ->0}, {}];
 
 opNNH = ExpandIndices[cNNH[gg1,gg2] anti[CC[NR[sp1,gg1]]].NR[sp1,gg2] Phibar[ii] Phi[ii]];
 
 opNNH + HC[opNNH]/.feynmangaugerules
]
\end{lstlisting}
For simplicity, we assume all WCs to be real. 
This implies that the number of independent real parameters 
associated with each non-hermitian operator in Tab.~\ref{tab:HiggsN} is half of that indicated in the column corresponding to $n_s=1$. 
Each effective coupling is characterised by new 
\texttt{InteractionOrder -> \{NP,1\}}.
The block \texttt{ONNH} corresponds to the first term of the 
dimension-five Lagrangian given in Eq.~\eqref{eq:L5}. 
The switch \texttt{FeynmanGauge} allows one to go from 
the Feynman gauge to the unitary gauge.
Having defined in \texttt{vSM.fr} the new physics scale $\Lambda$ as
\begin{lstlisting}
(* New physics scale *)
Lambda == {
 ParameterType    -> External, 
 BlockName        -> NEWPHYSICSSCALE,
 OrderBlock       -> 1,
 Value            -> 1000.0,
 ComplexParameter -> False,
 TeX              -> \[CapitalLambda],
 Description      -> "New physics scale [GeV]"
}
\end{lstlisting}
we obtain the dimension-five contribution to the full Lagrangian in Eq.~\eqref{eq:L} as
\begin{lstlisting}
(* Dimension-5 Lagrangian *)
LND5 := ONNH/Lambda
\end{lstlisting}

The procedure for implementing the dimension-six Lagrangian in Eq.~\eqref{eq:L6} is fully analogous to the one described above.
We just note that we also define the couplings $c_{NA}^{ij}$ and $c_{NZ}^{ij}$ 
as \textit{internal} parameters,  
using the expressions in Eq.~\eqref{eq:cNAcNZ}.

Further, in order to account for the shifts in $m_{N_1}$ and $y_\nu^{i1}$ 
induced by $\mathcal{O}_{NNH}^{11}$ and $\mathcal{O}_{LNH}^{i1}$, 
respectively, we perform ``finite renormalisation''. 
More specifically, upon electroweak symmetry breaking,
\begin{equation}
 \frac{c_{NNH}^{11}}{\Lambda} \mathcal{O}_{NNH}^{11} \to
 \frac{c_{NNH}^{11}v^2}{2\Lambda}\overline{N_{1R}^c}N_{1R} + \dots\,,
\end{equation}
and
\begin{equation}
 \frac{c_{LNH}^{i1}}{\Lambda^2} \mathcal{O}_{LNH}^{i1} \to
 \frac{c_{LNH}^{i1} v^2}{2 \Lambda^2} \frac{v}{\sqrt2}\overline{\nu_{iL}} N_{1R} + \dots\,,
\end{equation}
where we have used $H = (0, (v+h)/\sqrt2)^T$, 
and the dots stand for the interaction terms with the dynamical Higgs field $h$. 
We see that these operators correct the Majorana mass and the neutrino Yukawa couplings, so we redefine $m_{N_1}$ and $y_\nu^{i1}$ to absorb these shifts:
\begin{equation}
 m_{N_1} \to m_{N_1} + \frac{c_{NNH}^{11}v^2}{\Lambda}
 \qquad \text{and} \qquad
 y_\nu^{i1} \to y_\nu^{i1} + \frac{c_{LNH}^{i1} v^2}{2 \Lambda^2}\,.
\end{equation}
In this way, the WCs do not appear in the mass terms, 
but only in the neutrino interactions with $h$.
In particular, $m_{N_1}$ is the physical mass of $N_1$. 
In the model file, this finite renormalisation procedure 
is realised by adding to the 
Lagrangian the following pieces:
\begin{lstlisting}
LNHiggsD5 := Block[{mNNH,sp1,gg1,gg2},
    
    mNNH = ExpandIndices[-(1/2) (vev^2/Lambda) cNNH[gg1,gg2] anti[CC[NR[sp1,gg1]]].NR[sp1,gg2]];
    
    mNNH + HC[mNNH]
]
\end{lstlisting}
and
\begin{lstlisting}
LNHiggsD6 := Block[{yukLNH,sp1,ii,jj,ff,gg,feynmangaugerules},
    feynmangaugerules = If[Not[FeynmanGauge], {G0|GP|GPbar ->0}, {}];
    
    yukLNH = ExpandIndices[-(1/2) (vev^2/Lambda^2) cLNH[ff,gg] LLbar[sp1,ii,ff].NR[sp1,gg] Eps[ii, jj] Phibar[jj]];
    
    yukLNH + HC[yukLNH]/.feynmangaugerules
]
\end{lstlisting}

The full Lagrangian containing the two-fermion interactions with $N_R$ 
is given by
\begin{lstlisting}
LeffHiggsN:= LNSM + LNHiggsD5 + LNHiggsD6 + LND5 + LND6
\end{lstlisting}
where \texttt{LNSM} is the renormalisable part of the Lagrangian defined in 
\texttt{vSM.fr}, and \texttt{LND6} encodes the five dimension-six operators from 
Tab.~\ref{tab:HiggsN}:
\begin{lstlisting}
(* Dimension-6 Lagrangian *)
LND6 := (ONB + ONW + OHN + OHNe + OLNH)/Lambda^2
\end{lstlisting}

The \textsc{Mathematica} notebook \texttt{vSMEFT\_2-fermions\_Higgs-N.nb}
loads and merges the \textsc{FeynRules} models \texttt{SM\_v\_Majorana.fr}, 
\texttt{vSM.fr} and \texttt{vSMEFT\_2-fermions\_Higgs-N.fr}, performs a number of standard sanity checks (hermiticity of the full Lagrangian, normalisation of kinetic terms, mass spectrum) and computes Feynman rules. 
The corresponding UFO model is generated by evaluating
\begin{lstlisting}
FeynmanGauge = False
WriteUFO[LeffHiggsN]
\end{lstlisting}
The output is written to the folder \texttt{vSMEFT\_2-fermions\_Higgs-N\_UFO}, 
see Sec.~\ref{sec:HNUFO}.

\subsubsection{\label{sec:vSMEFT4fermionsN}\texttt{vSMEFT\_4-fermions\_single-N.fr}}
%
In this file, we implement the four-fermion operators with one RH neutrino field, 
see Tab.~\ref{tab:singleN}. Since each WC carries four flavour indices, 
but we are interested in the effective couplings 
to the first generation of RH neutrinos only, we fix the $N_R$ flavour index to 1 
and drop it from the WCs, \textit{e.g.} 
$c_{duNe}^{ijkl} \to c_{duNe}^{ij1l} \to c_{duNe}^{ijl}$. For this operator, we then define
\begin{lstlisting}
(* Wilson coefficients cduNe *)
cduNe == {
 ParameterType	  -> External,
 BlockName	  -> OPERATORduNe,
 Indices      -> {Index[Generation],Index[Generation],Index[Generation]},
 ComplexParameter -> False,
 Value	-> {cduNe[i_?NumericQ, j_?NumericQ, k_?NumericQ] -> 0.0},
 InteractionOrder -> {NP, 1},
 TeX		  -> Subscript[c, duNe],
 Description	  -> "Wilson coefficients cduNe"
}
\end{lstlisting}
and
\begin{lstlisting}
(* Operator OduNe *)
OduNe := Block[{opduNe,mu,sp1,sp2,sp3,sp4,cc,ff1,ff2,ff3},
  
 opduNe = ExpandIndices[cduNe[ff1,ff2,ff3] dRbar[sp1,ff1,cc].Ga[mu,sp1,sp2].uR[sp2,ff2,cc] NRbar[sp3,1].Ga[mu,sp3,sp4].lR[sp4,ff3]];
  
 opduNe + HC[opduNe]
]
\end{lstlisting}
Assuming the WCs are real, 
we have 27 real parameters $c_\mathcal{O}^{ijk}$ 
for each of the five single-$N_R$ operator structures 
in Tab.~\ref{tab:singleN}.

\subsubsection{\label{sec:vSMEFT4fermionsNN}\texttt{vSMEFT\_4-fermions\_pair-N.fr}}
%
The procedure in the case of pair-$N_R$ operators from Tab.~\ref{tab:pairN} is similar. 
Restricting ourselves to the case of effective couplings 
to the first generation of $N_R$ only, the WCs $c_{fN}^{ijkl}$ with $f=Q,u_R,d_R,L,e_R$, become $2\times2$ hermitian matrices $c_{fN}^{ij}$, 
whereas $c_{NN}$ is just one real number. Hence, in the \textsc{FeynRules} model file they are defined as follows:
\begin{lstlisting}
(* Wilson coefficient cNN *)
cNN == {
 ParameterType	  -> External,
 BlockName	  -> OPERATORNN, 
 OrderBlock	  -> 1,
 ComplexParameter -> False,
 Value		  -> 0.0,
 InteractionOrder -> {NP, 1},
 TeX		  -> Subscript[c, NN],
 Description	  -> "Wilson coefficient cNN"
}
\end{lstlisting}
and
\begin{lstlisting}
(* Wilson coefficients ceN *)
ceN == {
 ParameterType	  -> External,
 BlockName	  -> OPERATOReN,
 Indices	  -> {Index[Generation],Index[Generation]},
 ComplexParameter -> False,
 Hermitian	  -> True,
 Value		  -> {ceN[1,1] -> 0.0, ceN[1,2] -> 0.0, ceN[1,3] -> 0.0,
				       ceN[2,2] -> 0.0, ceN[2,3] -> 0.0,
                                                        ceN[3,3] -> 0.0},
 Definitions	  -> {ceN[i_?NumericQ,j_?NumericQ] :> ceN[j,i] /; i > j},
 InteractionOrder -> {NP, 1},
 TeX		  -> Subscript[c, eN],
 Description	  -> "Wilson coefficients ceN"
}
\end{lstlisting}
and similarly for $c_{LN}$, $c_{uN}$, $c_{dN}$ and $c_{QN}$. 
For simplicity, we assume the WCs to be real. In this case, each hermitian matrix 
$c_{fN}$ contains six real parameters $c_{fN}^{ij}$ with $i \leq j$.
The corresponding operators are defined as
\begin{lstlisting}
(* Operator ONN *)
ONN := Block[{mu,sp1,sp2,sp3,sp4},
  
 ExpandIndices[cNN NRbar[sp1,1].Ga[mu,sp1,sp2].NR[sp2,1] NRbar[sp3,1].Ga[mu,sp3,sp4].NR[sp4,1]]
  
]
\end{lstlisting}
and 
\begin{lstlisting}
(* Operator OeN *)
OeN := Block[{mu,sp1,sp2,sp3,sp4,ff1,ff2},

 ExpandIndices[ceN[ff1,ff2] lRbar[sp1,ff1].Ga[mu,sp1,sp2].lR[sp2,ff2] NRbar[sp3,1].Ga[mu,sp3,sp4].NR[sp4,1]]

]
\end{lstlisting}

However, a \textsc{UFO} model produced by direct export of four-fermion interactions 
containing a Majorana field 
is not supported by \textsc{MadGraph5\_aMC@NLO}. 
The reason is that fermion flow cannot be univocally defined in this case. 
We discuss how to overcome this issue in the next section.

\subsection{\label{sec:aux}Auxiliary-field implementation of four-fermion operators}
%
\textsc{UFO} models obtained by direct export of the single-$N_R$ and pair-$N_R$ 
four-fermion operators implemented in 
\texttt{\seqsplit{vSMEFT\_4-fermions\_single-N.fr}} and \texttt{vSMEFT\_4-fermions\_pair-N.fr}
cannot be run in \textsc{MadGraph5\_aMC@NLO}, 
because it \emph{cannot handle Majorana fermions 
in operators with more than two fermions}~\cite{Alwall:2014hca}. 
One possible workaround is to implement renormalisable 
UV completions for such operators containing new heavy 
scalar and/or vector fields, as discussed \textit{e.g.}~in Ref.~\cite{Cottin:2021lzz}. 
Another option is to implement Dirac $N$ in \textsc{FeynRules}/\textsc{UFO} 
and then make suitable replacements of effective couplings, 
as discussed in App.~B of Ref.~\cite{Bolton:2025tqw}. 
Here, we aim to provide a generic implementation of four-fermion operators 
with Majorana $N$, such that the end \textsc{MadGraph} user 
does not have to worry about the details of the \textsc{FeynRules} implementation.
We follow the strategy described below. 

We introduce auxiliary, non-propagating scalar or vector fields, 
replacing every four-fermion contact interaction 
by a sum of diagrams containing only three-point vertices. 
Each such vertex involves two fermion fields.
\begin{figure}
\label{fig:splitting}
\end{figure}

Consider a generic four-fermion operator $\mathcal{O} = J_1 J_2$ 
or $\mathcal{O} = J_1^\mu J_{2\mu}$, with $J_{1,2}$ ($J_{1,2}^\mu$) 
being scalar (vector) currents of the form $J = \overline{f}f'$ ($J^\mu = \overline{f}\gamma^\mu f')$, where $f$ and $f'$ are chiral fermion fields. 
The effective Lagrangian reads
\begin{equation}
 \mathcal{L}_{4f} = \frac{c}{\Lambda^2} \mathcal{O} + \text{h.c.} =  \frac{c}{\Lambda^2} J_1 J_2 + \text{h.c.}\,,
\end{equation}
where we have focused on the scalar current case 
and assumed that $\mathcal{O}$ is not hermitian.
We introduce two auxiliary, non-propagating complex scalar fields $U$ and $V$ 
with the following Lagrangian:
\begin{equation}
 \mathcal{L}_\mathrm{aux} = \Lambda^2 \left(U^\ast U - V^\ast V\right)
 + \frac{1}{\sqrt{2}}
 \left(U J_1 - c\, U^\ast J_2 + V J_1 + c\, V^\ast J_2 + \text{h.c.}\right)\,,
 \label{eq:Laux}
\end{equation}
where the first two terms should be viewed as formal quadratic terms 
rather than the physical mass terms.
The equations of motion (EoMs) for $U$ and $V$, 
$\partial\mathcal{L}_\mathrm{aux}/\partial U^\ast = 0$ and 
$\partial\mathcal{L}_\mathrm{aux}/\partial V^\ast = 0$, give
\begin{equation}
 U = -\frac{J_1^\dagger-c J_2}{\sqrt2 \Lambda^2}
 \qquad \text{and} \qquad
 V = \frac{J_1^\dagger+c J_2}{\sqrt2 \Lambda^2}\,.
 \label{eq:EoMs}
\end{equation}
The ``on-shell'' Lagrangian $\mathcal{L}_\mathrm{aux}$ evaluated at the EoMs 
reduces to $\mathcal{L}_{4f}$:
\begin{equation}
 \mathcal{L}_\mathrm{aux}|_\mathrm{EoMs} = \frac{c}{\Lambda^2} J_1 J_2 + \text{h.c.} = \mathcal{L}_{4f}\,.
\end{equation}
The two fields $U$ and $V$ with opposite-sign quadratic terms are needed to cancel 
the unwanted $J_1 J_1^\dagger$ and $J_2 J_2^\dagger$ terms, while preserving 
the desired $J_1 J_2$ and its hermitian conjugate. 
In the case of non-hermitian operators composed of two vector currents $J_1^\mu J_{2\mu}$, 
as \textit{e.g.} $\mathcal{O}_{duNe} = (\overline{d_{R}}\gamma^{\mu}u_{R}) (\overline{N_{R}}\gamma_{\mu}e_{R})$, 
the required auxiliary fields are complex vector fields $U^\mu$ and $V^\mu$.
 
The same procedure applies to an hermitian four-fermion operator, 
assuming real $U$ and $V$, and normalising their quadratic terms as $\Lambda^2/2$. 
Namely, in the case of vector currents, we have
\begin{equation}
 \mathcal{L}_\mathrm{aux} = \frac{\Lambda^2}{2}\left(U_\mu U^\mu-V_\mu V^\mu\right)
   +\frac{1}{\sqrt{2}}\left(U_\mu J_1^\mu-c\,U_\mu J_2^\mu+V_\mu J_1^\mu+c\,V_\mu J_2^\mu\right)\,.
 \label{eq:Laux2}
\end{equation}
The EoMs for $U$ and $V$ maintain the same form of Eq.~\eqref{eq:EoMs}. 
Note that in this case, $J_1^\dagger=J_1$ and  $J_2^\dagger=J_2$. 

We adopt the described procedure to implement the four-fermion interactions with Majorana $N$.

\subsubsection{\label{sec:vSMEFT4fermionsNaux}\texttt{vSMEFT\_4-fermions\_single-N\_aux.fr}}
%
We ``open up'' the single-$N_R$ operators introducing the auxiliary fields
summarised in Tab.~\ref{tab:auxN}. 
\begin{table}
 \renewcommand{\arraystretch}{1.75}
 \centering 
 \begin{tabular}[t]{|c|c|c|c|}
  \hline
  Operator & Auxiliary fields $X_i$ & $(\mathrm{SU}(3)_c,\mathrm{SU}(2)_L)_Y$ &
  Interactions with $X_i$  \\
  \hline
  $\mathcal{O}_{LNLe}^{ijk}$ &
  $U_{LNLe}^i\,,~V_{LNLe}^i$ &
  $(\1,\2)_{-1/2}$ & 
  $\overline{L_i}N_R X_i\,,~c_{LNLe}^{ijk} X_i^\dagger \eps \overline{L_j}^T e_{kR}$ \\
  \hline
  $\mathcal{O}_{duNe}^{ijk}$ &
 $U^{i,\mu}_{duNe}\,,~V^{i,\mu}_{duNe}$ &
  $(\1,\1)_{-1}$ & 
  $c_{duNe}^{ijk} \overline{d_{iR}}\gamma_\mu u_{jR} X_k^\mu\,,~
  \overline{N_R} \gamma_\mu e_{kR} X_k^{\mu\dagger} $ \\ 
 $\mathcal{O}_{LNQd}^{ijk}$ &
 $U_{LNLe}^i\,,~V_{LNLe}^i$ &
  $(\1,\2)_{-1/2}$ &
  $\overline{L_i}N_R X_i\,,~c_{LNQd}^{ijk} X_i^\dagger \eps \overline{Q_j}^T d_{kR}$ \\
  $\mathcal{O}_{LdQN}^{ijk}$ & 
  $U_{LdQN}^i\,,~V_{LdQN}^i$ &
  $(\3,\2)_{1/6}$ & 
  $c_{LdQN}^{ijk} X_k^\dagger \eps \overline{L_i} d_{jR}\,,~
  \overline{Q_k} N_R X_k$
    \\
 $\mathcal{O}_{QuNL}^{ijk}$ &
 $U_{QuNL}^i\,,~V_{QuNL}^i$ &
 $(\1,\2)_{-1/2}$ & 
 $c_{QuNL}^{ijk} \overline{Q_{iL}} u_{jR} X_k\,,~\overline{N_R} L_k X_k^\dagger$ \\
  \hline
 \end{tabular}
\caption{\label{tab:auxN}Auxiliary fields with their SM gauge quantum numbers and interactions for the four-fermion single-$N_R$ operators with a Majorana $N$.}
\end{table}
In the case of the operator $\mathcal{O}_{LNLe}$, 
the auxiliary fields $U_{LNLe}^{i,0}$ and $U_{LNLe}^{i,-}$ forming 
an SU(2)$_L$ doublet are defined 
in the model file \texttt{vSMEFT\_4-fermions\_single-N\_aux.fr} as
\begin{lstlisting}
(* Physical auxiliary fields *)
(* O_LNLe: U weak doublet, Y=-1/2. *)
S[201] == {
 ClassName	-> xLNLeU0, 
 ClassMembers	-> {xLNLeU01,xLNLeU02,xLNLeU03},
 Indices	-> {Index[Generation]}, 
 FlavorIndex	-> Generation, 
 SelfConjugate	-> False, 
 Mass		-> 0, 
 Width		-> 0,
 QuantumNumbers	-> {Q -> 0}, 
 PropagatorLabel-> {"XLU0","XLU01","XLU02","XLU03"}, 
 PropagatorType -> D,
 PropagatorArrow-> Forward, 
 PDG		-> {9010011,9010012,9010013},
 ParticleName	-> {"x_lnle_u0_1","x_lnle_u0_2","x_lnle_u0_3"},
 AntiParticleName-> {"x_lnle_u0_1~","x_lnle_u0_2~","x_lnle_u0_3~"}
},

S[202] == {
 ClassName	-> xLNLeUm,
 ClassMembers	-> {xLNLeUm1,xLNLeUm2,xLNLeUm3},
 Indices	-> {Index[Generation]}, 
 FlavorIndex	-> Generation, 
 SelfConjugate	-> False, 
 Mass		-> 0, 
 Width		-> 0,
 QuantumNumbers	-> {Q -> -1}, 
 PropagatorLabel-> {"XLUm","XLUm1","XLUm2","XLUm3"}, \
 PropagatorType	-> D,
 PropagatorArrow-> Forward, 
 PDG		-> {9010021,9010022,9010023},
 ParticleName	-> {"x_lnle_um_1","x_lnle_um_2","x_lnle_um_3"},
 AntiParticleName-> {"x_lnle_up_1","x_lnle_up_2","x_lnle_up_3"}
}
\end{lstlisting}
The masses here are placeholders. 
The UFO patch described in Sec.~\ref{sec:patch} replaces them by $\Lambda$ 
in the momentum-independent propagators.
The auxiliary fields $V_{LNLe}^{i,0}$ and $V_{LNLe}^{i,-}$ 
are defined in an analogous way. 
For convenience, we also define the doublet containers $U_{LNLe}^i$ and 
$V_{LNLe}^i$ used to write gauge-invariant interactions in a compact form. 
For instance,
\begin{lstlisting}
(* Unphysical auxiliary fields *)
(* O_LNLe: U weak doublet, Y=-1/2. *)
S[221] == {
 ClassName	-> xLNLeU, 
 Unphysical	-> True,
 Indices	-> {Index[SU2D],Index[Generation]}, 
 FlavorIndex	-> SU2D, 
 SelfConjugate	-> False,
 QuantumNumbers	-> {Y -> -1/2}, 
 Definitions	-> {xLNLeU[1,ff_] :> xLNLeU0[ff], xLNLeU[2,ff_] :> xLNLeUm[ff]}
}
\end{lstlisting}

Following Eq.~\eqref{eq:Laux}, the interaction Lagrangian $\mathcal{L}_\mathrm{aux}$ implementing $\mathcal{O}_{LNLe}$ reads
\begin{lstlisting}
LauxLNLe := Block[{bare,ii,jj,sp1,sp2,ff1,ff2,ff3},
 
 bare = ExpandIndices[(
      gAux4F xLNLeU[ii,ff1] LLbar[sp1,ii,ff1].NR[sp1,1]
    - xLNLeUbar[ii,ff1] cLNLe[ff1,ff2,ff3] Eps[ii,jj] LLbar[sp2,jj,ff2].lR[sp2,ff3]
    + gAux4F xLNLeV[ii,ff1] LLbar[sp1,ii,ff1].NR[sp1,1]
    + xLNLeVbar[ii,ff1] cLNLe[ff1,ff2,ff3] Eps[ii,jj] LLbar[sp2,jj,ff2].lR[sp2,ff3])/Sqrt[2]];
  
  bare + HC[bare]
]
\end{lstlisting}
where the bookkeeping coupling \texttt{gAux4F} has been defined in \texttt{vSM.fr} 
as follows: 
\begin{lstlisting}
gAux4F == {
 ParameterType	  -> Internal, 
 Value		  -> 1,
 InteractionOrder -> {AUX,1}, 
 TeX		  -> Subscript[g,aux],
 Description	  -> "Unit bookkeeping coupling 
 		      for non-propagating four-fermion auxiliaries"
} 
\end{lstlisting}

The implementation of the remaining four semi-leptonic single-$N_R$ operators is fully analogous 
to the one of $\mathcal{O}_{LNLe}$. The final auxiliary Lagrangian 
for the four-fermion single-$N_R$ operators is
\begin{lstlisting}
LN4fermions1Aux := LauxLNLe + LauxduNe + LauxLNQd + LauxLdQN + LauxQuNL
\end{lstlisting}
and the full Lagrangian is
\begin{lstlisting}
Leff4fermions1Aux := LNSM + LN4fermions1Aux
\end{lstlisting}
A \textsc{UFO} model obtained by exporting this Lagrangian is free of any issue related to the definition of fermion flow and, 
after running it through the \textsc{UFO} patch described in Sec.~\ref{sec:patch},
can be readily used in \textsc{MadGraph5\_aMC@NLO}.

\subsubsection{\label{sec:vSMEFT4fermionsNNaux}\texttt{vSMEFT\_4-fermions\_pair-N\_aux.fr}}
%
The auxiliary fields for the pair-$N_R$ operators are summarised in Tab.~\ref{tab:auxNN}.
\begin{table}
 \renewcommand{\arraystretch}{1.75}
 \centering 
 \begin{tabular}[t]{|c|c|c|c|}
  \hline
  Operator & Auxiliary fields $X_i$ & $(\mathrm{SU}(3)_c,\mathrm{SU}(2)_L)_Y$ &
  Interactions with $X_i$  \\
  \hline
  $\mathcal{O}_{NN}^{ij}$ &
  $U_{NN}^\mu\,,~V_{NN}^\mu$ &
  $(\1,\1)_{0}$ & 
  $\overline{N_R} \gamma_\mu N_R X^\mu\,,~
  c_{NN} \overline{N_R} \gamma_\mu N_R X^\mu$ \\
  \hline
  $\mathcal{O}_{f N}^{ij}$ &
  $U_{fN}^\mu\,,~V_{fN}^\mu$ &
  $(\1,\1)_{0}$ & 
  $c_{fN}^{ij} \overline{f_i} \gamma_\mu f_j X^\mu\,,~
  \overline{N_R} \gamma_\mu N_R X^\mu$ \\
  \hline
 \end{tabular}
\caption{\label{tab:auxNN}Auxiliary fields with their SM gauge quantum numbers and interactions for the four-fermion pair-$N_R$ operators with a Majorana $N$. Here $f$ stands for a SM fermion gauge multiplet, $f = Q, u_R, d_R, L, e_R$.}
\end{table}
All of them are SM neutral real vector fields. 
For instance, for the operator $\mathcal{O}_{dN}$, we have
\begin{lstlisting}
(* Physical auxiliary fields *)
(* OdN: U and V real vectors *)
V[309] == {
 ClassName	 -> xdNU, 
 SelfConjugate	 -> True, 
 Mass		 -> 0, 
 Width		 -> 0,
 PropagatorLabel -> "XdN_U", 
 PropagatorType	 -> Wavy,
 PDG		 -> 9030091, 
 ParticleName	 -> "x_pair_dn_u0"
},

V[310] == {
 ClassName	 -> xdNV, 
 SelfConjugate	 -> True, 
 Mass		 -> 0, 
 Width		 -> 0,
 PropagatorLabel -> "XdN_V", 
 PropagatorType	 -> Wavy,
 PDG		 -> 9030101, 
 ParticleName	 -> "x_pair_dn_v0"
}
\end{lstlisting}
Again, the \textsc{UFO} patch replaces the mass value by $\Lambda$ in the momentum-independent propagators. 

Following Eq.~\eqref{eq:Laux2}, the auxiliary Lagrangian $\mathcal{L}_\mathrm{aux}$ 
implementing $\mathcal{O}_{dN}$ is given by
\begin{lstlisting}
LauxdN := Block[{mu,sp1,sp2,sp3,sp4,cc,ff1,ff2},
 
 ExpandIndices[(
      gAux4F xdNU[mu] NRbar[sp3,1].Ga[mu,sp3,sp4].NR[sp4,1]
    - xdNU[mu] cdN[ff1,ff2] dRbar[sp1,ff1,cc].Ga[mu,sp1,sp2].dR[sp2,ff2,cc]
    + gAux4F xdNV[mu] NRbar[sp3,1].Ga[mu,sp3,sp4].NR[sp4,1]
    + xdNV[mu] cdN[ff1,ff2] dRbar[sp1,ff1,cc].Ga[mu,sp1,sp2].dR[sp2,ff2,cc]
  )/Sqrt[2]]
]

\end{lstlisting}
The implementation of the remaining five pair-$N_R$ operators proceeds in the same manner. 
The final auxiliary Lagrangian for the four-fermion pair-$N_R$ operators is
\begin{lstlisting}
LN4fermions2Aux := LauxNN + LauxeN + LauxLN + LauxuN + LauxdN + LauxQN
\end{lstlisting}
and the full Lagrangian is 
\begin{lstlisting}
Leff4fermions2Aux := LNSM + LN4fermions2Aux
\end{lstlisting}
This Lagrangian leads to a \textsc{UFO} 
model for the pair-$N_R$ operators with a Majorana $N$ compatible with \textsc{MadGraph5\_aMC@NLO}.

\subsubsection{\label{sec:vSMEFT}\texttt{vSMEFT.fr}}
%
Finally, we combine in \texttt{vSMEFT.fr} all the effective operators 
from Tabs.~\ref{tab:HiggsN}--\ref{tab:pairN} 
(assuming effective couplings to the first generation of $N_R$ only). 
The full $\nu$SMEFT Lagrangian up to dimension six is given by
\begin{lstlisting}
LvSMEFT:= LeffHiggsN + LN4fermions1Aux + LN4fermions2Aux
\end{lstlisting}
While this full Lagrangian is rather lengthy, 
one would need to use this model if interested 
in an interplay between the two- and four-fermion operators with $N_R$.

\subsection{\label{sec:UFO}\textsc{UFO} models}
%
%
\subsubsection{\label{sec:HNUFO}\texttt{vSMEFT\_2-fermions\_Higgs-N\_UFO}}
%
The folder \texttt{vSMEFT\_2-fermions\_Higgs-N\_UFO} containing the \textsc{UFO} model should be added 
to the folder \texttt{models} of the \textsc{MadGraph5\_aMC@NLO} directory. 
The $\nu$SMEFT blocks in the parameter card of the model
corresponding to the new physics scale $\Lambda$ (in GeV)
and to the WCs of the six Higgs-$N_R$ operators (see~Tab.~\ref{tab:HiggsN}) are
\begin{lstlisting}
###################################
## INFORMATION FOR NEWPHYSICSSCALE
###################################
Block newphysicsscale 
    1 1.000000e+03 # Lambda 

###################################
## INFORMATION FOR OPERATORHN
###################################
Block operatorhn 
    1   1 0.000000e+00 # cHN1x1 

###################################
## INFORMATION FOR OPERATORHNE
###################################
Block operatorhne 
    1   1 0.000000e+00 # cHNe1x1 
    1   2 0.000000e+00 # cHNe1x2 
    1   3 0.000000e+00 # cHNe1x3 

###################################
## INFORMATION FOR OPERATORLNH
###################################
Block operatorlnh 
    1   1 0.000000e+00 # cLNH1x1 
    2   1 0.000000e+00 # cLNH2x1 
    3   1 0.000000e+00 # cLNH3x1 

###################################
## INFORMATION FOR OPERATORNB
###################################
Block operatornb 
    1   1 0.000000e+00 # cNB1x1 
    2   1 0.000000e+00 # cNB2x1 
    3   1 0.000000e+00 # cNB3x1  

###################################
## INFORMATION FOR OPERATORNNH
###################################
Block operatornnh 
    1   1 0.000000e+00 # cNNH1x1 

###################################
## INFORMATION FOR OPERATORNW
###################################
Block operatornw 
    1   1 0.000000e+00 # cNW1x1 
    2   1 0.000000e+00 # cNW2x1 
    3   1 0.000000e+00 # cNW3x1 
\end{lstlisting}
By default, all WCs are set to zero. 
This is also the case for the active-heavy mixing matrix elements $V_{\alpha N_j}$ with $\alpha = e,\mu,\tau$ and $j=1,2,3$. 
They are defined as in the \href{https://feynrules.irmp.ucl.ac.be/wiki/HeavyN}{\texttt{HeavyN}} model.
\begin{lstlisting}
###################################
## INFORMATION FOR NUMIXING
###################################
Block numixing 
    1 0.000000e+00 # VeN1 
    2 0.000000e+00 # VeN2 
    3 0.000000e+00 # VeN3 
    4 0.000000e+00 # VmuN1 
    5 0.000000e+00 # VmuN2 
    6 0.000000e+00 # VmuN3 
    7 0.000000e+00 # VtaN1 
    8 0.000000e+00 # VtaN2 
    9 0.000000e+00 # VtaN3
\end{lstlisting}
%

\subsubsection{\label{sec:patch}\textsc{UFO} patch}
%
The file \texttt{patch\_vSMEFT\_aux\_ufo.py} 
is a script that should be run after every fresh \texttt{WriteUFO} export 
of \texttt{Leff4fermions1Aux} and/or \texttt{Leff4fermions2Aux}:
\begin{lstlisting}
python3 /path/to/patch_vSMEFT_aux_ufo.py /path/to/vSMEFT_4-fermions_[*]-N_aux_UFO
\end{lstlisting}
where \texttt{[*]} should be replaced by either \texttt{single} or \texttt{pair}.

The patch modifies the files 
\texttt{object\_library.py}, \texttt{particles.py} and \texttt{propagators.py}
contained in the corresponding \textsc{UFO} folder. 
Namely, it performs the following actions:
\begin{enumerate}
 \item Adds momentum-independent propagators $+i/\Lambda^2$ and $-i/\Lambda^2$ for scalar auxiliary fields $U$ and $V$, respectively; 
 and $+i g_{\mu\nu}/\Lambda^2$ and $-i g_{\mu\nu}/\Lambda^2$ for vector auxiliary fields $U_\mu$ and $V_\mu$, respectively.
 \item Assigns \texttt{`mass = Param.Lambda'} to auxiliary particles solely so \texttt{`Mass(id)\textasciicircum2'} supplies the propagator normalisation.
 \item Marks auxiliary particles \texttt{`propagating = False'}, preventing them from being treated as physical external states.
 \item Preserves the custom propagator when the \textsc{UFO} constructs antiparticles.
 \item Creates \texttt{`*.pre\_aux\_patch'} back-ups and can be run repeatedly.
\end{enumerate}

We would like to stress that the auxiliary fields $U$ and $V$ 
are \textit{not} finite-mass UV mediators. Their propagators have opposite signs 
and no momentum-dependent pole.

\subsubsection{\label{sec:1NUFO}\texttt{vSMEFT\_4-fermions\_single-N\_aux\_UFO}}
%
The $\nu$SMEFT block in the parameter card of the model
corresponding to the WCs of the operator $\mathcal{O}_{duNe}$ is
\begin{lstlisting}
###################################
## INFORMATION FOR OPERATORDUNE
###################################
Block operatordune 
    1   1   1 0.000000e+00 # cduNe1x1x1 
    1   1   2 0.000000e+00 # cduNe1x1x2 
    1   1   3 0.000000e+00 # cduNe1x1x3 
    1   2   1 0.000000e+00 # cduNe1x2x1 
    1   2   2 0.000000e+00 # cduNe1x2x2 
    1   2   3 0.000000e+00 # cduNe1x2x3 
    1   3   1 0.000000e+00 # cduNe1x3x1 
    1   3   2 0.000000e+00 # cduNe1x3x2 
    1   3   3 0.000000e+00 # cduNe1x3x3 
    2   1   1 0.000000e+00 # cduNe2x1x1 
    2   1   2 0.000000e+00 # cduNe2x1x2 
    2   1   3 0.000000e+00 # cduNe2x1x3 
    2   2   1 0.000000e+00 # cduNe2x2x1 
    2   2   2 0.000000e+00 # cduNe2x2x2 
    2   2   3 0.000000e+00 # cduNe2x2x3 
    2   3   1 0.000000e+00 # cduNe2x3x1 
    2   3   2 0.000000e+00 # cduNe2x3x2 
    2   3   3 0.000000e+00 # cduNe2x3x3 
    3   1   1 0.000000e+00 # cduNe3x1x1 
    3   1   2 0.000000e+00 # cduNe3x1x2 
    3   1   3 0.000000e+00 # cduNe3x1x3 
    3   2   1 0.000000e+00 # cduNe3x2x1 
    3   2   2 0.000000e+00 # cduNe3x2x2 
    3   2   3 0.000000e+00 # cduNe3x2x3 
    3   3   1 0.000000e+00 # cduNe3x3x1 
    3   3   2 0.000000e+00 # cduNe3x3x2 
    3   3   3 0.000000e+00 # cduNe3x3x3
\end{lstlisting}
Similar blocks are present for the remaining four single-$N_R$ operators from Tab.~\ref{tab:singleN}. 
In total, there are $27 \times 5 = 135$ real WCs in the model file.

The masses of the auxiliary fields \texttt{x\_dune\_um\_i} and \texttt{x\_dune\_vm\_i}, 
with \texttt{i = 1,2,3}, as well as of the other 48 auxiliary fields, are set to \texttt{Lambda} by the \textsc{UFO} patch.

\subsubsection{\label{sec:2NUFO}\texttt{vSMEFT\_4-fermions\_pair-N\_aux\_UFO}}
%
The $\nu$SMEFT blocks in the parameter card of the model
corresponding to the WCs of the six pair-$N_R$ 
four-fermion operators (see~Tab.~\ref{tab:pairN}) are
\begin{lstlisting}
###################################
## INFORMATION FOR OPERATORDN
###################################
Block operatordn 
    1   1 0.000000e+00 # cdN1x1 
    1   2 0.000000e+00 # cdN1x2 
    1   3 0.000000e+00 # cdN1x3 
    2   2 0.000000e+00 # cdN2x2 
    2   3 0.000000e+00 # cdN2x3 
    3   3 0.000000e+00 # cdN3x3 

###################################
## INFORMATION FOR OPERATOREN
###################################
Block operatoren 
    1   1 0.000000e+00 # ceN1x1 
    1   2 0.000000e+00 # ceN1x2 
    1   3 0.000000e+00 # ceN1x3 
    2   2 0.000000e+00 # ceN2x2 
    2   3 0.000000e+00 # ceN2x3 
    3   3 0.000000e+00 # ceN3x3 

###################################
## INFORMATION FOR OPERATORLN
###################################
Block operatorln 
    1   1 0.000000e+00 # cLN1x1 
    1   2 0.000000e+00 # cLN1x2 
    1   3 0.000000e+00 # cLN1x3 
    2   2 0.000000e+00 # cLN2x2 
    2   3 0.000000e+00 # cLN2x3 
    3   3 0.000000e+00 # cLN3x3 

###################################
## INFORMATION FOR OPERATORNN
###################################
Block operatornn 
    1 0.000000e+00 # cNN 

###################################
## INFORMATION FOR OPERATORQN
###################################
Block operatorqn 
    1   1 0.000000e+00 # cQN1x1 
    1   2 0.000000e+00 # cQN1x2 
    1   3 0.000000e+00 # cQN1x3 
    2   2 0.000000e+00 # cQN2x2 
    2   3 0.000000e+00 # cQN2x3 
    3   3 0.000000e+00 # cQN3x3 

###################################
## INFORMATION FOR OPERATORUN
###################################
Block operatorun 
    1   1 0.000000e+00 # cuN1x1 
    1   2 0.000000e+00 # cuN1x2 
    1   3 0.000000e+00 # cuN1x3 
    2   2 0.000000e+00 # cuN2x2 
    2   3 0.000000e+00 # cuN2x3 
    3   3 0.000000e+00 # cuN3x3 
\end{lstlisting}

The masses of the auxiliary fields \texttt{x\_pair\_[name]\_u0} and \texttt{x\_pair\_[name]\_v0}, with \texttt{[name] = nn, en, ln, un, dn, qn}, are set to \texttt{Lambda} by the \textsc{UFO} patch.

\subsubsection{\texttt{vSMEFT\_UFO}}
%
This is the \textsc{UFO} model containing all the considered effective interactions with $N_R$. It is obtained from \texttt{vSMEFT.fr} by executing
\begin{lstlisting}
FeynmanGauge = False
WriteUFO[LvSMEFT]
\end{lstlisting}
in the \textsc{Mathematica} notebook \texttt{vSMEFT.nb} 
and patching the obtained folder with the \textsc{UFO} patch 
described in Sec.~\ref{sec:patch}.

\section{\label{sec:validation}Model validation}
The initial version of the \texttt{vSMEFT\_2-fermions\_Higgs-N\_UFO} model was largely validated in Ref.~\cite{Butterworth:2019iff}, 
where the focus was on the Higgs phenomenology at the LHC 
in the presence of a relatively light $N_1$
with $0.01~\mathrm{GeV} \lesssim m_{N_1} \lesssim 10~\mathrm{GeV}$. 
In this mass range, the dominant decay channel is $N_1 \to \nu \gamma  $ 
induced by $\mathcal{O}_{NA}^{i1}$~\cite{Duarte:2015iba}.
It was assumed that $c_{HN}^{11} = c_{HNe}^{1i} = c_{NZ}^{i1} = 0$ 
to avoid stringent bounds from $Z \to \nu\nu\gamma\gamma$, 
modification of the $W$-width and $Z \to \nu\nu\gamma$, respectively.
The remaining set of operators, namely, $\mathcal{O}_{NA}^{i1}$, 
$\mathcal{O}_{LNH}^{i1}$ and $\mathcal{O}_{NNH}^{11}$, 
leads to the following processes, that were investigated in Ref.~\cite{Butterworth:2019iff}:
\begin{itemize}
\item $pp \to \gamma^\ast \to \nu \nu \gamma$ through $\mathcal{O}_{NA}^{i1}$, 
meaning $pp \to \gamma^\ast \to \nu N_1$, $N_1 \to \nu \gamma$;
\item $pp \to W^\pm \to e_i^\pm \nu \gamma$ through $\mathcal{O}_{NA}^{i1}$, 
meaning $pp \to W^\pm \to e_i^\pm N_1$, $N_1 \to \nu \gamma$;
\item $pp \to h \to \nu \nu \gamma$ through $\mathcal{O}_{LNH}^{i1}$, meaning 
$pp \to h \to \nu N_1$, $N_1 \to \nu \gamma$;
\item $pp \to h \to \nu \nu \gamma \gamma$ through $\mathcal{O}_{NA}^{i1}$, 
meaning $pp \to h \to \nu N_1 \gamma$, $N_1 \to \nu \gamma$;
\item $pp \to h \to \nu \nu \gamma \gamma$ through $\mathcal{O}_{NNH}^{11}$,
meaning $pp \to h \to N_1 N_1$, $N_1 \to \nu \gamma$.
\end{itemize}
%

Below we consider some representative processes 
triggered by other operators, namely, 
$\mathcal{O}_{HN}^{11}$ and $\mathcal{O}_{HNe}^{1i}$, 
and compare the obtained results with those existing in the literature. 
These operators lead to new neutral current and charged current 
interactions of $N_1$. 
Upon electroweak symmetry breaking, we have%
\footnote{Note that, for simplicity, we do not distinguish here between the flavour and mass eigenstates of sterile neutrinos. In other words,
we assume the mixing matrix elements $V_{sN_j} \approx \delta_{sN_j}$, 
with $s$ denoting the sterile neutrino flavour.}
\begin{align}
 \frac{c_{HN}^{11}}{\Lambda^2}\mathcal{O}_{HN}^{11} &\to -\frac{g}{c_w} \frac{c_{HN}^{11} v^2}{2 \Lambda^2}\, 
 \overline{N_{1R}} \gamma^\mu N_{1R}\, Z_\mu + \dots\,, \\
 \frac{c_{HNe}^{1i}}{\Lambda^2}\mathcal{O}_{HNe}^{1i} &\to \frac{g}{\sqrt2} 
 \frac{c_{HNe}^{1i} v^2}{2 \Lambda^2}\, \overline{N_{1R}} \gamma^\mu e_{iR}\, W_\mu^+ + \dots\,,
\end{align}
where $g$ is the SU(2)$_L$ gauge coupling.

The examples provided below have a two-fold scope: 
(i)~validate the model 
and
(ii)~show a beginner how to use it.

\begin{figure}[t]
\centering
\includegraphics[width=\textwidth]{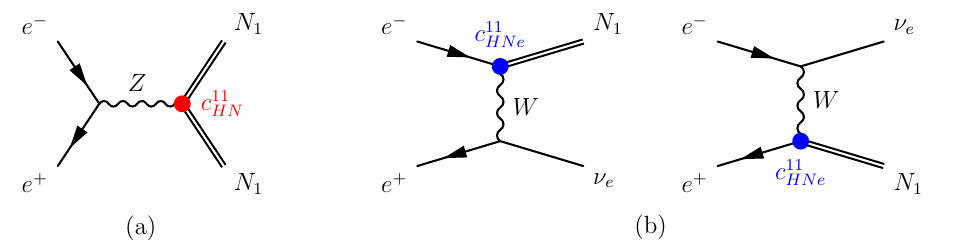}
\caption{Pair-$N_1$ and single-$N_1$ production at 
a LEP-like $e^+e^-$ collider 
through modified (a) neutral and 
(b) charged current interactions, respectively.}
\label{fig:diagsee}
\end{figure}

First example is pair-$N_1$ and single-$N_1$ production 
at a LEP-like $e^+ e^-$ collider, namely,
\begin{itemize}
 \item $e^+ e^- \to Z \to N_1 N_1$ through $\mathcal{O}_{HN}^{11}$, 
 see Fig.~\ref{fig:diagsee}(a);
 \item $e^+ e^- \to \nu_e N_1$ through $\mathcal{O}_{HNe}^{11}$, 
 see Fig.~\ref{fig:diagsee}(b).
\end{itemize}
Such processes have been considered \textit{e.g.}~in Ref.~\cite{Bolton:2025tqw}. 
Therein, in App.~B, the effective couplings 
\begin{equation}
 [Z_N^R]^{11} \equiv \frac{c_{HN}^{11} v^2}{2\Lambda^2} 
 \qquad \text{and} \qquad
 [W_N^R]^{11} \equiv \frac{c_{HNe}^{11} v^2}{2\Lambda^2}
\end{equation}
were set to $10^{-2}$ (one at a time). 
Assuming $\Lambda = 1$~TeV, this translates to 
$c_{HN}^{11} \approx 0.33$ and $c_{HNe}^{11} \approx 0.33$.

Let us calculate the cross sections for these processes 
at the $Z$-pole, as a function of $m_{N_1}$, with our \textsc{UFO} model 
\texttt{vSMEFT\_2-fermions\_Higgs-N\_UFO}. 
After launching \textsc{MadGraph5\_aMC@NLO}, for the first process,
one needs to execute the following commands:
\begin{lstlisting}
MG5_aMC> import model vSMEFT_2-fermions_Higgs-N_UFO
MG5_aMC> generate e+ e- > n1 n1
MG5_aMC> output vSMEFT/ee_to_NN
MG5_aMC> launch
> set mN1 scan:[0,10,20,30,40,41,42,43,44,45]
> set cHN1x1 0.33
\end{lstlisting}
indicating the desired beam IDs ($e^+e^-$ collider, hence \texttt{No PDF}) 
and energy, $\sqrt{s} = 91.2$~GeV, in the run card:
\begin{lstlisting}
#*********************************************************************
# Collider type and energy                                           *
# lpp: 0=No PDF, 1=proton, -1=antiproton, 2=photon from proton,      *
#                3=photon from electron, 4=photon from muon          *
#*********************************************************************
     0        = lpp1    ! beam 1 type 
     0        = lpp2    ! beam 2 type
     45.6     = ebeam1  ! beam 1 total energy in GeV
     45.6     = ebeam2  ! beam 2 total energy in GeV
\end{lstlisting}

Similarly, the second process can be simulated by
\begin{lstlisting}
MG5_aMC> generate e+ e- > ve n1
MG5_aMC> output vSMEFT/ee_to_vN
MG5_aMC> launch
> set mN1 scan:[0,20,40,60,70,75,80,85,88,90]
> set cHNe1x1 0.33
\end{lstlisting}

\begin{figure}[t]
\centering
\includegraphics[width=0.49\textwidth]{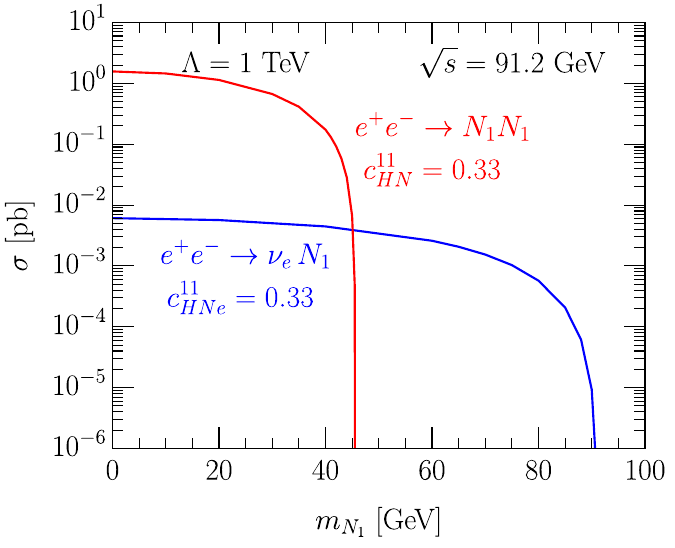}
\hfill
\includegraphics[width=0.49\textwidth]{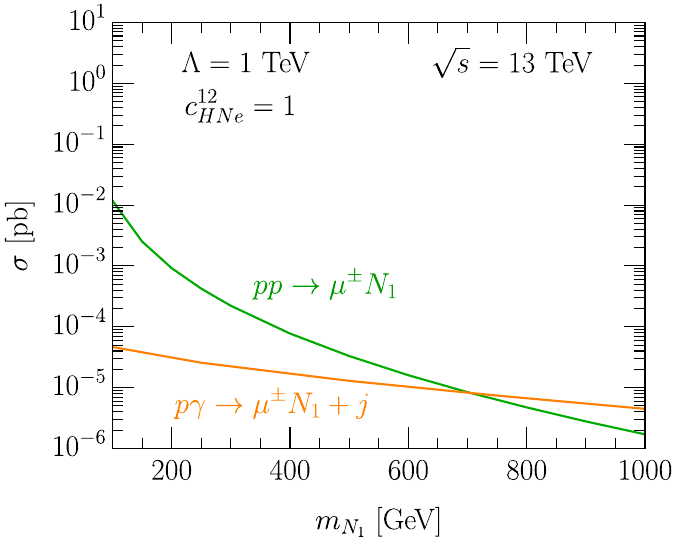}
\caption{\textit{Left panel}: cross sections for $e^+e^- \to N_1 N_1$ 
triggered by $c_{HN}^{11}/\Lambda^2 = 0.33$~TeV$^{-2}$, 
and $e^+e^- \to \nu_e N_1$ 
triggered by $c_{HNe}^{11}/\Lambda^2 = 0.33$~TeV$^{-2}$, 
as a function of $m_{N_1}$, at a LEP-like collider with $\sqrt{s} = 91.2$~GeV. 
\textit{Right panel}: cross sections for 
$pp \to \mu^\pm N_1$ and $p\gamma \to \mu^\pm N_1 + j$ 
triggered by $c_{HNe}^{12}/\Lambda^2 = 1$~TeV$^{-2}$, 
as a function of $m_{N_1}$, at the LHC with $\sqrt{s} = 13$~TeV.}
\label{fig:Xsecs}
\end{figure}

The resulting cross sections are shown in the left panel of 
Fig.~\ref{fig:Xsecs}. They are in perfect agreement with the corresponding 
lines in Figs.~17 and 16 (right panel) of Ref.~\cite{Bolton:2025tqw}.

\begin{figure}[t]
\centering
\includegraphics[width=\textwidth]{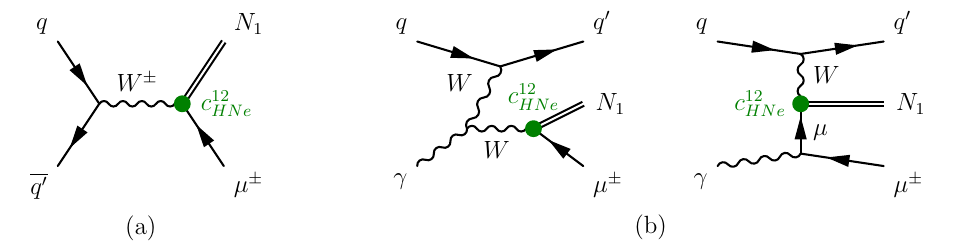}
\caption{Single-$N_1$ production at the LHC
via (a) RH charged current Drell-Yan process, and
(b) $W\gamma$ fusion (representative diagrams).}
\label{fig:diagspp}
\end{figure}

The second example is single-$N_1$ production via 
the RH charged current interaction at the LHC:
\begin{itemize}
 \item $pp \to W^{\pm*} \to \mu^\pm N_1$ through $\mathcal{O}_{HNe}^{12}$, see Fig.~\ref{fig:diagspp}(a);
 \item $p\gamma \to \mu^\pm N_1 + j$ through $\mathcal{O}_{HNe}^{12}$, 
 see Fig.~\ref{fig:diagspp}(b).
\end{itemize}
Such processes have been studied in Ref.~\cite{Cirigliano:2021peb} 
focusing on a heavier $N_1$ with $m_W < m_{N_1} < \Lambda$.

The first process is calculated by 
\begin{lstlisting}
MG5_aMC> define mu = mu+ mu-
MG5_aMC> generate p p > mu n1
MG5_aMC> output vSMEFT/pp_to_muN
MG5_aMC> launch
> set mN1 scan:[100,150,200,250,300,400,500,600,700,800,900,1000]
> set cHNe1x2 1
\end{lstlisting}
setting the beam IDs corresponding to protons and energy, $\sqrt{s} = 13$~TeV, in the run card:
\begin{lstlisting}
#*********************************************************************
# Collider type and energy                                           *
# lpp: 0=No PDF, 1=proton, -1=antiproton, 2=photon from proton,      *
#                3=photon from electron, 4=photon from muon          *
#*********************************************************************
     1        = lpp1    ! beam 1 type 
     1        = lpp2    ! beam 2 type
     6500.0   = ebeam1  ! beam 1 total energy in GeV
     6500.0   = ebeam2  ! beam 2 total energy in GeV
\end{lstlisting}

The second process is simulated by
\begin{lstlisting}
MG5_aMC> generate p a > mu n1 j
MG5_aMC> add process a p > mu n1 j
MG5_aMC> output vSMEFT/pa_to_muNj
MG5_aMC> launch
> set mN1 scan:[100,250,500,750,1000]
> set cHNe1x2 1
\end{lstlisting}
In the run card, we keep the beam IDs to their default values, 
\textit{i.e.}~\texttt{1 = lpp1} and \texttt{1~=~lpp2}, as suggested in Ref.~\cite{Pascoli:2018heg}.

The cross sections for both processes are shown in the right panel 
of Fig.~\ref{fig:Xsecs}. They approximately reproduce those presented 
in the left panel of Fig.~4 in Ref.~\cite{Cirigliano:2021peb}, 
where these processes have been computed 
at next-to-leading order (NLO) in QCD.

It is instructive to check the implementation of four-fermion operators 
based on the auxiliary fields against analytical calculations. Let's consider 
the process $d \bar{d} \to N_1 N_1$ triggered by $\mathcal{O}_{dN}^{11}$. 
In the case of Majorana $N_1$, the cross section is
\begin{equation}
 \sigma(d \bar{d} \to N_1 N_1) = \frac{c_{dN}^2}{144 \pi \Lambda^4} 
 s \left(1-\frac{4 m_{N}^2}{s}\right)^{3/2}\,,
\end{equation}
whereas if $N_1$ were a Dirac fermion, the cross section would be 
\begin{equation}
 \sigma(d \bar{d} \to N_1 \bar{N}_1) = \frac{c_{dN}^2}{192 \pi \Lambda^4}
  s \sqrt{1-\frac{4 m_{N}^2}{s}}\left[1+ \frac{1}{3}\left(1-\frac{4 m_{N}^2}{s}\right)
    \right]\,,
\end{equation}
see \textit{e.g.} Ref.~\cite{Cottin:2021lzz}.

We simulate this process in \textsc{MadGraph5\_aMC@NLO} 
with our \textsc{UFO} model \linebreak \texttt{vSMEFT\_4-fermions\_pair-N\_aux\_UFO} 
by
\begin{lstlisting}
MG5_aMC> import model vSMEFT_4-fermions_pair-N_aux_UFO
MG5_aMC> generate d d~ > n1 n1
MG5_aMC> output vSMEFT/ddbar_to_NN
MG5_aMC> launch
> set mN1 scan:[0,50,100,150,200,250,300,350,400,450]
> set cdN1x1 1
\end{lstlisting}
setting in the run card
\begin{lstlisting}
#*********************************************************************
# Collider type and energy                                           *
# lpp: 0=No PDF, 1=proton, -1=antiproton,                            *
#                2=elastic photon of proton/ion beam                 *
#             +/-3=PDF of electron/positron beam                     *
#             +/-4=PDF of muon/antimuon beam                         *
#*********************************************************************
     0         = lpp1    ! beam 1 type 
     0         = lpp2    ! beam 2 type
     500.0     = ebeam1  ! beam 1 total energy in GeV
     500.0     = ebeam2  ! beam 2 total energy in GeV
\end{lstlisting}
corresponding to $\sqrt{s} = 1$~TeV. 
Next, we also perform a scan over the beam energy \texttt{ebeam = ebeam1 = ebeam2} in the range from 150~GeV to 500~GeV setting $m_{N_1} = 100$~GeV.

\begin{figure}[t]
\centering
\includegraphics[width=0.49\textwidth]{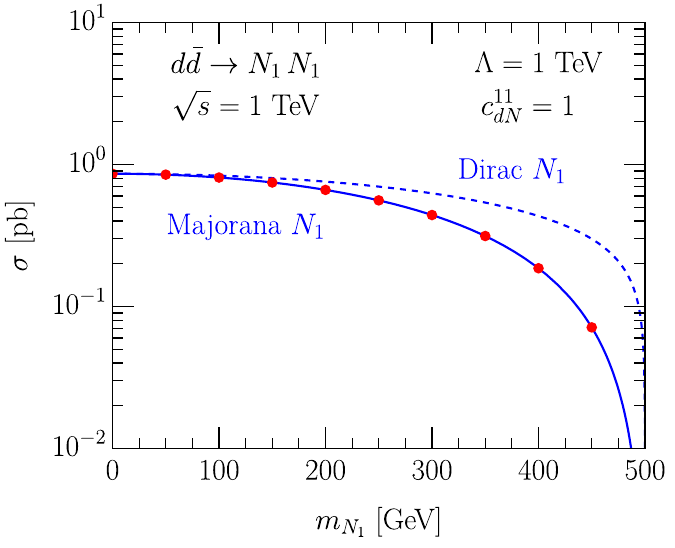}
\hfill
\includegraphics[width=0.49\textwidth]{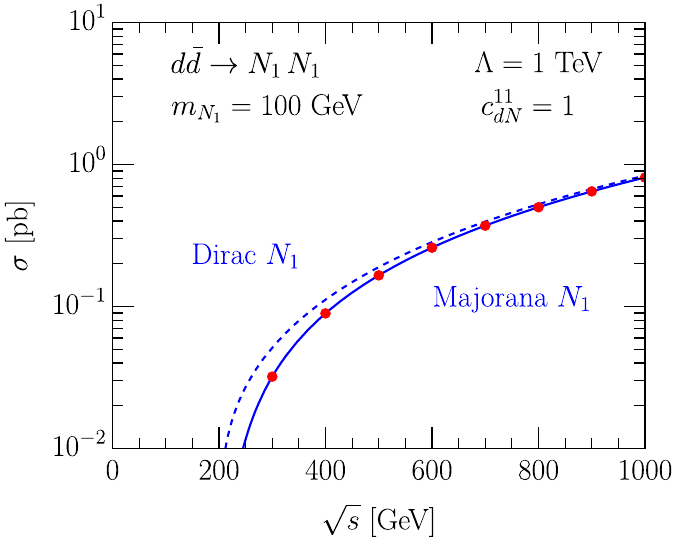}
\caption{\textit{Left panel}: cross section for $d \bar{d} \to N_1 N_1$ 
triggered by $c_{dN}^{11}/\Lambda^2 = 1$~TeV$^{-2}$
as a function of $m_{N_1}$ for $\sqrt{s} = 1$~TeV 
calculated analytically (\textit{solid line}) and with \textsc{MadGraph5\_aMC@NLO} 
(\textit{filled circles}). The \textit{dashed line} is the analytical result for the cross section 
of $d \bar{d} \to N_1 \bar{N}_1$ in the case of Dirac $N_1$. 
\textit{Right panel}: the same cross section, but as a function of $\sqrt{s}$ for $m_{N_1} = 100$~GeV.}
\label{fig:Xsecs2}
\end{figure}

In the left panel of Fig.~\ref{fig:Xsecs2}, we show the analytical result (solid line) 
and the result of the simulation (filled circles) for $\sigma(d\bar{d} \to N_1 N_1)$ 
as a function of $m_{N_1}$,
setting $c_{dN}^{11}=1$ and $\Lambda = \sqrt{s} = 1$~TeV. 
For comparison, we also display the analytical result in the case of Dirac $N_1$. 
In the right panel, we show the same cross section, but as a function of $\sqrt{s}$ 
assuming $m_{N_1} = 100$~GeV. 
We find perfect agreement between the analytical and numerical results.
%

\section{\label{sec:outlook}Summary and outlook}
In the presence of (i) sterile neutrinos with masses below or around 
the electroweak scale, $v$, and 
(ii) new heavy physics at a scale $\Lambda \gg v$, 
the most general description of new physics effects at low energies, 
$E \lesssim \Lambda$, is in terms of the $\nu$SMEFT. 
In this note, we focused on the dimension-five and dimension-six 
operators involving RH neutrinos. 
We implemented them in the \textsc{FeynRules} model files 
\texttt{vSMEFT\_2-fermions\_Higgs-N.fr}, 
\texttt{vSMEFT\_4-fermions\_single-N.fr} and 
\texttt{vSMEFT\_4-fermions\_pair-N.fr}. 

The former file was used to generate the \textsc{UFO} model 
\texttt{vSMEFT\_2-fermions\_Higgs-N\_UFO}, that in turn allows one 
to perform simulations with \textsc{MadGraph5\_aMC@NLO}. 
However, the UFO models obtained by direct export of 
the four-fermion operators with a Majorana $N$ implemented in 
the latter \textsc{FeynRules} files are not supported by \textsc{MadGraph5\_aMC@NLO}, because fermion flow cannot be 
unambiguously defined in this case.

In view of this, we developed an implementation of such four-fermion operators 
relying on auxiliary, non-propagating fields. 
The auxiliary fields allow one to replace every four-fermion contact interaction by a sum of diagrams containing three-point vertices only, avoiding the issue with the definition of fermion flow, see Sec.~\ref{sec:aux}. The \textsc{FeynRules} model files 
\texttt{vSMEFT\_4-fermions\_single-N\_aux.fr} and 
\texttt{vSMEFT\_4-fermions\_pair-N\_aux.fr} \linebreak 
along with the provided \textsc{UFO} patch, see Sec.~\ref{sec:patch}, 
were used to generate the \textsc{UFO} models 
\texttt{vSMEFT\_4-fermions\_single-N\_aux\_UFO} and
\texttt{vSMEFT\_4-fermions\_pair-N\_aux\_UFO} compatible with \textsc{MadGraph5\_aMC@NLO}. 
Finally, all the effective interactions of interest are combined in a single \textsc{UFO} 
model \texttt{vSMEFT\_UFO}.

The current version of the models is suitable for 
performing tree-level computations 
(with the loop-induced Higgs-gluon-gluon vertex implemented as 
an effective coupling). 
As a next step, the models could further be adapted for computations at NLO in QCD, 
similarly to the \href{https://feynrules.irmp.ucl.ac.be/wiki/HeavyN}{\texttt{HeavyN}} model. Further, one could also combine the developed models 
with the \href{https://smeftsim.github.io}{\texttt{SMEFTsim}} package 
to study an interplay between the effective operators with and without RH neutrinos.

Both \textsc{FeynRules} and \textsc{UFO} 
models are made publicly available on 
\href{https://github.com/arsenii-titov/vSMEFT.git}{\textsc{GitHub}~\faGithub}. 
We hope that the released model files 
will be useful not only for further phenomenological studies, 
but also to experimental collaborations, in particular, at the LHC, 
for implementing and performing specific analyses 
in the framework of the $\nu$SMEFT.

\small
\section*{Acknowledgements}
I would like to thank 
J.\,M.~Butterworth, M.~Chala, C.~Englert and M.~Spannowsky for the collaboration on the phenomenology of the Higgs-$N_R$ operators in Ref.~\cite{Butterworth:2019iff}, 
and R.~Beltrán, G.~Cottin, J.\,C.~Helo, M.~Hirsch and Z.\,S.~Wang 
for the collaboration on the phenomenology of the four-fermion operators with $N_R$ in Refs.~\cite{Cottin:2021lzz,Beltran:2021hpq}.
I am indebted to C.~Hays and M.~Spannowsky 
for encouraging the public release of the model files. 
I am further grateful to
M.~Chala, M.~Hirsch, Z.\,S.~Wang, R.~Beltrán and G.~Cottin   
for the fruitful collaboration 
on various aspects of the $\nu$SMEFT over the past few years. 
I also acknowledge discussions with R.~Ruiz on the 
\href{https://feynrules.irmp.ucl.ac.be/wiki/HeavyN}{\texttt{HeavyN}} model.
This work has been funded by the European Union, NextGenerationEU, 
National Recovery and Resilience Plan 
(mission~4, component~2) 
under the project \textit{MODIPAC: Modular Invariance in Particle Physics and Cosmology} (CUP~C93C24004940006). 
The auxiliary-field implementation of four-fermion operators 
with a Majorana $N$ was 
developed with the help of Codex (OpenAI).
Feynman diagrams were drawn using \texttt{feynMF}/\texttt{feynMP}~\cite{Ohl:1995kr}.

\footnotesize
\bibliography{vSMEFT.bib}

\end{document}